\documentclass{article}
\pdfoutput=1
\usepackage[T1]{fontenc}
\usepackage[latin9]{inputenc}
\usepackage{amsmath}
\usepackage{amssymb}
\usepackage{graphicx}
\usepackage{subfigure}
\usepackage{CJK}
\usepackage{amsmath}
\usepackage{cases}
\usepackage{indentfirst}
\usepackage{cite}

\begin{document}
\title{Statistical Mechanics of the Directed 2-distance Minimal Dominating Set problem}
\author{$Yusupjan\quad Habibulla$ \\School of Physics and Technology, Xinjiang University, Sheng-Li Road 666,\\  Urumqi 830046, China}

\maketitle
\begin{abstract}
The directed L-distance minimal dominating set (MDS) problem has wide practical applications in the fields of computer science and communication networks. Here, we study this problem from the perspective of purely theoretical interest. We only give results for an Erd$\acute{o}$s R$\acute{e}$nyi (ER) random graph and regular random graph, but this work can be extended to any type of networks. We develop spin glass theory to study the directed 2-distance MDS problem. First, we find that the belief propagation algorithm does not converge when the inverse temperature exceeds a threshold on either an ER random network or regular random network. Second, the entropy density of replica symmetric theory has a transition point at a finite inverse temperature on a regular random graph when the node degree exceeds 4 and on an ER random graph when the node degree exceeds 6.6; there is no entropy transition point (or $\beta=\infty$) in other circumstances. Third, the results of the replica symmetry (RS) theory are in perfect agreement with those of belief propagation (BP) algorithm while the results of the belief propagation decimation (BPD) algorithm are better than those of the greedy heuristic algorithm. \\\\
\textbf{\large Keywords: }Directed 2-distance Minimal Dominating Set, Belief Propagation, Erd$\acute{o}$s R$\acute{e}$nyi Random Graph, Regular Random Graph, Belief Propagation Decimation.
\end{abstract}
\section{ Introduction}
There is a close relationship among the minimal dominating set (MDS), the directed minimal dominating set (DMDS), the L-distance minimal dominating set (L-MDS) and the directed L-distance minimal dominating set (D-LMDS). The idea behind the MDS is to construct the node set of smallest size such that any node of the network is either in this set or is adjacent to at least one node of this set. A DMDS is the smallest set of nodes such that each node either belongs to this set or has at least one parent node in this set. The allocation of network resources that satisfies a given service with the least use of resources is a frequent problem in the field of communication networks. We are interested in the problem of computing the minimum set of occupied nodes (referred to henceforth as servers) such that every node is occupied or has at least one occupied parent node at a distance of at most $L$ (i.e., the L-MDS problem), where the distance between two nodes in the graph is the minimum number of hops necessary to go from one node to the other. Each parent server then provides a service to or monitors those child nodes within a distance $L$. Consider a simple network $W$ formed by $N$ nodes and $M$ arcs (directed edges), with each arc $(i, j)$ pointing from a parent node (predecessor) $i$ to a child node (successor) $j$. The arc density $\alpha$ is defined simply as $\alpha \equiv M/N$, and the mean arc density $C$ is defined as $C\equiv 2\times\alpha$. There is a set $\gamma $; if any node of the network belongs to this set or at least one parent neighbor node within 2-distance belongs to this set, then this set $\gamma$ is called D-2MDS of the given network W. If node $i$ belongs to D-2MDS, we say $i$ is occupied. In Fig. 1, the green nodes (i.e., occupied nodes) construct D-2MDS for the given small graph. If node $i$ belongs to D-2MDS or at least one parent node within 2-distance is occupied, we then say $i$ is observed. In the figure, each blue node has at least one occupied parent neighbor node within 2-distance so that it is observed.\\
The L-MDS problem arises mainly in the design of a communication network in the real world. L-MDS has important applications in several fields; e.g., communication networks [1], locating servers [2] and copying a distributed directory [3]. There are various types of L-MDS problem, such as Liar's dominating set [4], the extended dominating set [5], the 2-distance paired dominating set [6], the [1,2]-dominating set [7], the ($\sigma , \rho $) dominating set [8], and the k-tuple dominating set [9]. The DMDS has wide application in the field of biological networks, such as for the spread of infectious disease [10], genetic regulation [11,12] and chemical reactions and metabolic regulation [13], and it is also applied in social, information and neuroscience fields [14-19]. If we naturally extend the regular DMDS problem, we get the corresponding D-2MDS problem; e.g., it is a regular DMDS problem in power generation and transportation [20]. If we assume that a power station can transport power to the 2-distance neighbor, then this power supply problem converts to a D-2MDS problem. There have been few works on the D-2MDS problem, and they only consider special cases; e.g., Wang and Chang [21] studied the unique minimum distance dominating set in directed split-stars using distributed algorithms while Banerjee et al. [22] introduced the directed d-hop MDS in which the in-degree equals 1. \\   
\begin{figure}[!htbp]
\centering
\includegraphics[width=10cm,height=10cm]{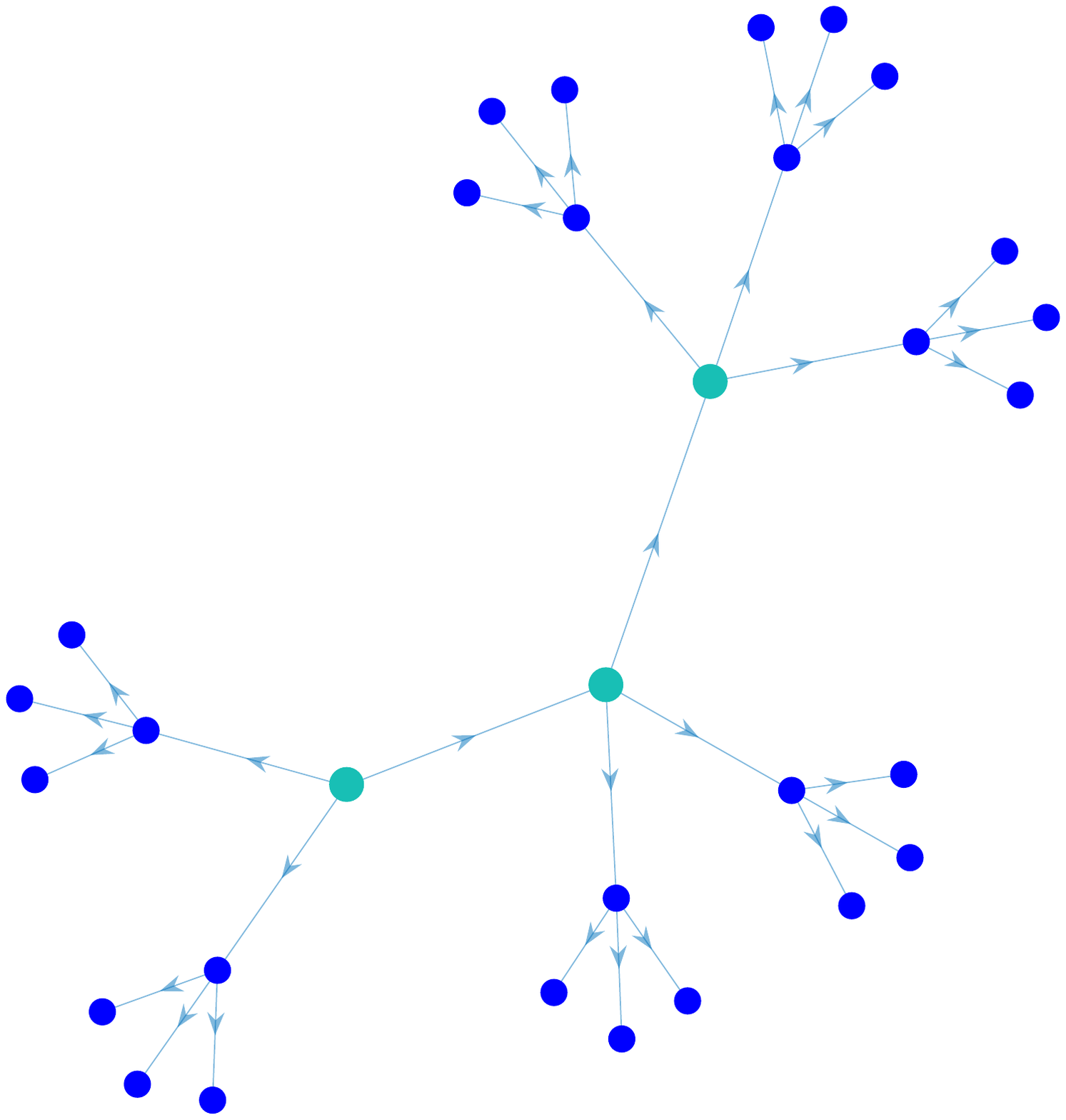}
\caption{Example of the directed 2-distance minimum dominating set. A small network with N = 31 nodes and M = 30 arcs. Green indicates an occupied node while blue indicates an empty but observed node. The three occupied nodes form a D-2MDS for this network.}
\end{figure} 
We adopt spin glass theory in this paper. Spin glass theory has wide application for optimization problems, such as k-sat [23], vertex cover [24], the feedback vertex set problem [25] and dominating set problem [26,27]. We recently used spin glass theory to study the regular MDS problem [28-31]. We introduced belief propagation decimation (BPD), warning propagation and survey propagation decimation algorithms to get the MDS and found that our algorithms rapidly provided solutions close to the optimal solution. This year, we used spin glass theory to study the undirected 2-distance MDS problem [32] and developed a BPD algorithm and greedy algorithm to estimate the size of the 2-distance MDS. In the present paper, we adopt statistical physics to study the D-2MDS, finding that the entropy density of the directed network work like with undirected network. This indicates that the solution spaces of the directed and undirected networks are connected with each other. We will study the solution space using one-step replica-symmetry (RS) breaking theory and compare our results with more rigorous mathematical results on the special network. We use three algorithms, namely population dynamics, BPD and greedy heuristic algorithms, to calculate the D-2MDS and find that the results of population dynamics and BPD algorithms are consistently better than those of the greedy heuristic algorithm on both Erd$\acute{o}$s R$\acute{e}$nyi (ER) random networks and regular random (RR) networks.\\
The remainder of the paper is organized as follows. Section 2 introduces RS theory for the D-2MDS problem and presents the belief propagation (BP) equation and corresponding thermodynamic quantities. Section 3 introduces the BPD and greedy algorithm for the D-2MDS problem, deriving the BP equation and marginal probability equation for the different node state condition. Section 4 summarizes our results and presents conclusions. 
\section{RS}
This section introduces mean field theory for the D-2MDS problem. The partition function plays an important role, with the entire calculation starting from the partition function. However, deriving the correct partition function is not a simple task. We assume that each node only interacts with its neighbors at the same time, so each node has a constraint on all its neighbor nodes. Depending on the RS mean field theory of the statistical physics, we write the partition function Z as
\begin{equation}
Z=\sum_{\underline{c}}\prod_{i\in W}{e^{-\beta \delta_{ c_{i}}^{0}}}\{1-(1-\delta_{ c_{i}}^{0})\prod_{j\in\partial i^{+}}(1- \delta_{ c_{j}}^{c_{i}-1})-\Theta[(\sum\limits_{j\in\partial i^{+}}\delta_{ c_{j}}^{c_{i}-2}+\sum\limits_{j\in\partial i^{-}}\delta_{ c_{j}}^{c_{i}+2})-1]\},
\end{equation}
where the Kronecker symbol $\delta_{m}^{n}=1$ if $m = n$ and $\delta_{m}^{n}=0$ otherwise while the function $\Theta(x)=1$ if $x\geq 0$ and $\Theta(x)=0$ otherwise. The symbol $\underline{ c}\equiv( c_{1}, c_{2},......, c_{n})$ denotes one of the $3^{n}$ configurations. $\beta$ is the inverse temperature. $\partial i^{+}$ denotes all the predecessors of node $i$ while $\partial i^{-}$ denotes all the successors of node $i$. The partition function therefore only takes into account all the directed 2-distance dominating set (D-2DS), and at $\beta\rightarrow\infty$ it is contributed exclusively by the D-2MDS configurations.\\
RS mean field theories, such as the Bethe--Peierls approximation [28] and partition function expansion [29], give a good estimation of the above spin glass model. These two theories provide the same results. In the present work, we derive the BP equation using the Bethe--Peierls approximation theory. We define on each arc $(i, j)$ of digraph D a distribution function $p_{i\rightarrow j}^{(c_{i},c_{j})}$ , which is the probability of node $i$ being in state $c_{i}$ and node $j$ being in state $c_{j}$ if the constraint of $j$ is not considered, and another distribution function $p_{j\leftarrow i}^{(c_{j},c_{i})}$, which is the probability of $j$ being in state $c_{j}$ and $i$ being in state $c_{i}$ if the constraint of $i$ is not considered. These messages must satisfy the equations

\begin{equation}
p_{i\leftarrow j}^{(c_{i},c_{j})}=\frac{e^{-\beta \delta_{ c_{i}}^{0}}A_{i\leftarrow j}^{(c_{i},c_{j})}}{\sum\limits_{\acute{ c}_{i},\acute{ c}_{j}}e^{-\beta \delta_{\acute c_{i}}^{0}}A_{i\leftarrow j}^{(\acute c_{i},\acute c_{j})}},
\end{equation}

\begin{equation}
p_{i\rightarrow j}^{(c_{i},c_{j})}=\frac{e^{-\beta \delta_{ c_{i}}^{0}}A_{i\rightarrow j}^{(c_{i},c_{j})}}{\sum\limits_{\acute{ c}_{i},\acute{ c}_{j}}e^{-\beta \delta_{\acute c_{i}}^{0}}A_{i\rightarrow j}^{(\acute c_{i},\acute c_{j})}},
\end{equation}

\begin{equation}
A_{i\leftarrow j}^{(c_{i},c_{j})}=[\prod\limits_{k\in\partial i^{+}\backslash j}\sum\limits_{c_{k}\in A^{+}}p_{k\rightarrow i}^{( c_{k}, c_{i})}-R_{i\leftarrow j}^{(c_{i},c_{j})}\prod\limits_{k\in\partial i^{+}\backslash j}\sum\limits_{c_{k}\geq c_{i}}p_{k\rightarrow i}^{(c_{k},c_{i})}]\prod\limits_{k\in\partial i^{-}}\sum\limits_{c_{k}\in A^{-}}p_{k\leftarrow i}^{( c_{k}, c_{i})},
\end{equation}

\begin{equation}
R_{i\leftarrow j}^{(c_{i},c_{j})}=(1-\delta_{ c_{i}}^{0})(\delta_{ c_{j}}^{c_{i}}+\delta_{ c_{j}}^{c_{i}+1}),
\end{equation}

\begin{equation}
A_{i\rightarrow j}^{(c_{i},c_{j})}=[\prod\limits_{k\in\partial i^{+}}\sum\limits_{c_{k}\in A^{+}}p_{k\rightarrow i}^{( c_{k}, c_{i})}-(1-\delta_{ c_{i}}^{0})\prod\limits_{k\in\partial i^{+}}\sum\limits_{c_{k}\geq c_{i}}p_{k\rightarrow i}^{(c_{k},c_{i})}]\prod\limits_{k\in\partial i^{-}\backslash j}\sum\limits_{c_{k}\in A^{-}}p_{k\leftarrow i}^{( c_{k}, c_{i})}.
\end{equation}

These two equations (2) and (3) are collectively referred to as the BP equation. The cavity message $p_{i\rightarrow j}^{(c_{i},c_{j})}$ represents the joint probability that the predecessor $i$ is in occupation state $c_{i}$ and its adjacent successor $j$ is in occupation state $c_{j}$ when the constraint of node $j$ is not considered. The cavity message $p_{i\leftarrow j}^{(c_{i},c_{j})}$ represents the joint probability that the successor node $i$ is in occupation state $c_{i}$ and its adjacent predecessor $j$ is in occupation state $c_{j}$ when the constraint of node $j$ is not considered. If node $i$ is in state $c_{i}=0$, it requests the successors only take state $c_{k}=0$ or $c_{k}=1$, and state $c_{k}=2$ is forbidden, but the predecessors can take any state. If node $i$ is in state $c_{i}=1$, it allows neighbor nodes to take any state, but at least one predecessor must be occupied. If node $i$ is in state $c_{i}=2$, it requests the predecessors only take state $c_{k}=1$ or $c_{k}=2$, at least one predecessor must be in state $c_{i}=1$, state $c_{k}=0$ is forbidden for the predecessors, and successors can take any state. $A^{+}$ represents the set of possible predecessor states while $A^{-}$ represents the set of possible successor states. The marginal probability $p_{i}^{c}$ of node $i$ is expressed as
\begin{equation}
p_{i}^{c_{i}}=\frac{e^{-\beta \delta_{ c_{i}}^{0}}A_{i}^{c_{i}}}{\sum\limits_{\acute{ c}_{i}}e^{-\beta \delta_{\acute c_{i}}^{0}}A_{i}^{\acute c_{i}}},
\end{equation}

\begin{equation}
A_{i}^{c_{i}}=[\prod\limits_{k\in\partial i^{+}}\sum\limits_{c_{k}\in A^{+}}p_{k\rightarrow i}^{( c_{k}, c_{i})}-(1-\delta_{ c_{i}}^{0})\prod\limits_{k\in\partial i^{+}}\sum\limits_{c_{k}\geq c_{i}}p_{k\rightarrow i}^{(c_{k},c_{i})}]\prod\limits_{k\in\partial i^{-}}\sum\limits_{c_{k}\in A^{-}}p_{k\leftarrow i}^{( c_{k}, c_{i})}.
\end{equation}

We calculate the marginal probability using the converged cavity messages $p_{k\rightarrow i}^{( c_{k}, c_{i})}$ and $p_{k\leftarrow i}^{( c_{k}, c_{i})}$. $p_{i}^{0}$ denotes the probability that node $i$ is occupied, $p_{i}^{1}$ denotes the probability that node $i$ has at least one occupied parent neighbor, and $p_{i}^{2}$ denotes the probability that node $i$ has at least one occupied 2-distance parent neighbor.

The free energy is finally calculated using mean field theory as
\begin{equation}
F_{0}=\sum_{i=1}^{N}F_{i}-\sum_{(i,j)\in G} F_{(i,j)},
\end{equation}
where

\begin{equation}\label{eq:12}
\begin{split}
F_{i}=-&\frac{1}{\beta}\ln[\sum\limits_{c_{i}}e^{-\beta \delta_{c_{i}}^{0}}[\prod\limits_{k\in\partial i^{+}}\sum\limits_{c_{k}\in A^{+}}p_{k\rightarrow i}^{( c_{k}, c_{i})}-(1-\delta_{ c_{i}}^{0})\prod\limits_{k\in\partial i^{+}}\sum\limits_{c_{k}\geq c_{i}}p_{k\rightarrow i}^{(c_{k},c_{i})}]\\
&\times\prod\limits_{k\in\partial i^{-}}\sum\limits_{c_{k}\in A^{-}}p_{k\leftarrow i}^{( c_{k}, c_{i})}]
\end{split},
\end{equation}

\begin{equation}
F_{(i,j)}=-\frac{1}{\beta}\ln[\sum_{ c_{i}, c_{j}}p_{i\rightarrow j}^{( c_{i}, c_{j})}p_{j\leftarrow i}^{( c_{j}, c_{i})}].
\end{equation}

Here, $F_{i}$ denotes the free energy of function node $i$ and $F_{(i,j)}$ denotes the free energy of the edge $(i,j)$. We iterate the BP equation until it converges to one stable point and then calculate the energy density $E=1/N\sum_{i}p_{i}^{0}$ and the mean free energy $f\equiv F/N$ using equations (7) and (9). The entropy density is calculated as $s=\beta(E-f)$.

\begin{figure}[htb]
  \centering
  \includegraphics[width=12cm,height=7cm]{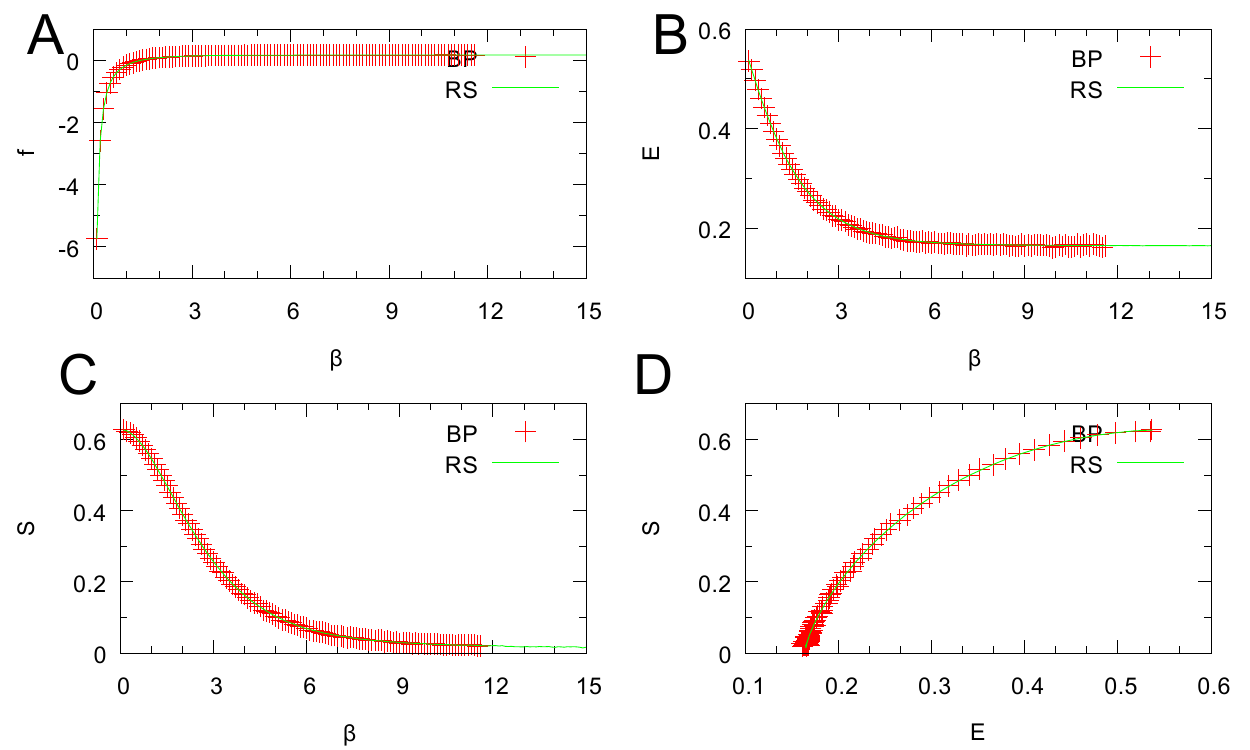}
  \caption{RS and BP results for the D-2MDS problem on the ER random graph with mean arc density $c=5$ and $N=10,000$ obtained using the BP equation and population dynamics. In subgraphs A, B, and C, the $x$-axis denotes the inverse temperature $\beta $ while the $y$-axis denotes the thermodynamic quantities. In subgraph D, the $x$-axis denotes the energy density while the $y$-axis denotes the entropy density.}
\end{figure}

We find that when the inverse temperature $\beta$ is larger than certain threshold value, BP iteration is unable to converge to a fixed point. Figure 2 shows that the BP equation does not converge when $\beta > 11.6$ on the ER random graph that mean arc density equals five. The entropy density is always positive, and with an increase in inverse temperature, the change rate of entropy becomes smaller and smaller, such that the entropy density reaches the transition point when the inverse temperature is extremely high. We use population dynamics to calculate the ground-state energy of the ER random graph. The energy density reaches a stable oscillation when the inverse temperature exceeds 10 while the entropy density is very close to zero, and we therefore use the average value of the energy density when the inverse temperature exceeds 10 to determine the ground-state energy.

\begin{table}[!htbp]
\caption{Inverse temperature $\beta_{d}$ at the entropy transition point and corresponding energy density for the ER random graph.}
\begin{tabular}{p{0.9cm}p{0.9cm}p{0.9cm}p{0.9cm}p{0.9cm}p{0.9cm}p{0.9cm}p{0.9cm}p{0.9cm}}
\hline
C & 6.7 & 7 & 8 & 9 & 10 & 11 & 12 &13\\
\hline
$\beta_{d}\approx$ & 14.6 & 13.15 & 11.05 & 10.25&10.05&10.04&10.17&10.3\\
\hline
$E_{min}\approx$ & 0.1059 & 0.0989 & 0.0799 & 0.0661&0.0555&0.0475&0.0414&0.0364\\
\hline
\end{tabular}
\end{table}

\begin{figure}[!htb]
  \centering
  \includegraphics[width=12cm,height=7cm]{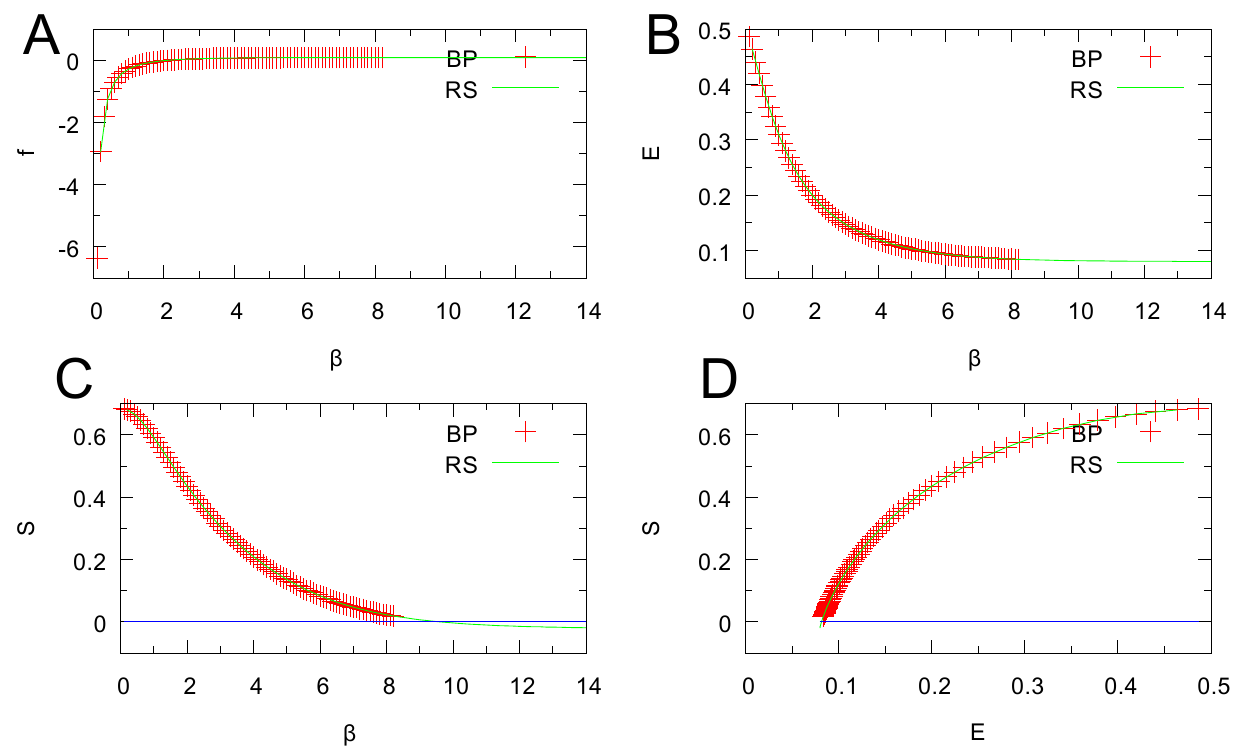}
  \caption{RS and BP results for the D-2MDS problem on the RR random graph with mean arc density $c=7$ and $N=10^{4}$ obtained using the BP equation and population dynamics. In subgraphs A, B, and C, the $x$-axis denotes the inverse temperature $\beta $ while the $y$-axis denotes the thermodynamic quantity. In subgraph D, the $x$-axis denotes the energy density while the $y$-axis denotes the entropy density.}
\end{figure}

Figure 3 shows that the population dynamics equation has a transition point of the entropy density on the RR graph when the mean arc density $C=5$ while the BP iteration can not converge when the $\beta >8.2$, and we thus did not need to average over the population dynamics results in this range. Tables 1 and 2 show that the phase transition temperature $\beta_{d}$ first decreases and then increases with mean arc density for both ER random and RR networks, but the energy density is always a monotonic decreasing function of mean arc density for these two networks. 
\begin{table}[!htb]
\caption{Inverse temperature $\beta_{d}$ at the entropy transition point and corresponding energy density for an RR graph.}
\begin{tabular}{p{0.9cm}p{0.9cm}p{0.9cm}p{0.9cm}p{0.9cm}p{0.9cm}p{0.9cm}p{0.9cm}p{0.9cm}}
\hline
C & 5 & 6 & 7 & 8 & 9 & 10 & 11 & 12\\
\hline
$\beta_{d}\approx$ & 19 & 10.6 & 9.9 & 9.4&9.5&9.8&10&10.3\\
\hline
$E_{min}\approx$ & 0.1255 & 0.0995 & 0.0813 & 0.0684&0.0585&0.0509&0.0448&0.0398 \\
\hline
\end{tabular}
\end{table}

\section{BPD Algorithm and Greedy Algorithm}
In this work, we use two algorithms to construct the solution of the given graph; i.e., the BPD algorithm and greedy algorithm. The greedy algorithm is fast but does not guarantee good results. The BPD algorithm is not as fast but always provides a good estimation for the 2-distance MDS problem.  
\subsection{BPD Algorithm}
If a node $i$ is unobserved (i.e., it is empty and no parent neighbor or 2-distance predecessors are occupied), the cavity message $p_{i\rightarrow j}$ on the arc $(i\rightarrow j)$ and the cavity message $p_{i\leftarrow j}$ on the arc $(i\leftarrow j)$ between nodes $i$ and $j$ are updated according to Eqs. (2) and (3). By contrast, if node $i$ is empty but observed and it has at least one occupied parent neighbor node, namely $c_{i}=1$, then this node does not restrict the states of any of its unoccupied parent neighbors. For such a node $i$, there is no opportunity to take $c_{i}=2$, and the cavity message $p_{i\rightarrow j}$ or $p_{i\leftarrow j}$ on link $(i,j)$ is then updated according to the equations

\begin{equation}
p_{i\leftarrow j}^{(c_{i},c_{j})}=\frac{e^{-\beta \delta_{ c_{i}}^{0}}(1-\delta_{c_{i}}^{2})\prod\limits_{k\in\partial i^{+}\backslash j}\sum\limits_{c_{k}\in A^{+}}p_{k\rightarrow i}^{( c_{k}, c_{i})}\prod\limits_{k\in\partial i^{-}}\sum\limits_{c_{k}\in A^{-}}p_{k\rightarrow i}^{( c_{k}, c_{i})}}{\sum\limits_{\acute{ c}_{i},\acute{ c}_{j}}e^{-\beta \delta_{\acute c_{i}}^{0}}(1-\delta_{c_{i}}^{2})[\prod\limits_{k\in\partial i^{+}\backslash j}\sum\limits_{\acute c_{k}\in A^{+}}p_{k\rightarrow i}^{( \acute c_{k}, \acute c_{i})}\prod\limits_{k\in\partial i^{-}}\sum\limits_{\acute c_{k}\in A^{-}}p_{k\rightarrow i}^{( \acute c_{k}, \acute c_{i})}},
\end{equation}

\begin{equation}
p_{i\rightarrow j}^{(c_{i},c_{j})}=\frac{e^{-\beta \delta_{ c_{i}}^{0}}(1-\delta_{c_{i}}^{2})\prod\limits_{k\in\partial i^{+}}\sum\limits_{c_{k}\in A^{+}}p_{k\rightarrow i}^{( c_{k}, c_{i})}\prod\limits_{k\in\partial i^{-}\backslash j}\sum\limits_{c_{k}\in A^{-}}p_{k\rightarrow i}^{( c_{k}, c_{i})}}{\sum\limits_{\acute{ c}_{i},\acute{ c}_{j}}e^{-\beta \delta_{\acute c_{i}}^{0}}(1-\delta_{c_{i}}^{2})\prod\limits_{k\in\partial i^{+}}\sum\limits_{\acute c_{k}\in A^{+}}p_{k\rightarrow i}^{( \acute c_{k}, \acute c_{i})}\prod\limits_{k\in\partial i^{-}\backslash j}\sum\limits_{\acute c_{k}\in A^{-}}p_{k\rightarrow i}^{( \acute c_{k}, \acute c_{i})}}.
\end{equation}

For node $i (c_{i}=1)$, at least one parent neighbor node $j$ is occupied and a message is sent to node $i$ as $p_{j\rightarrow i}^{(0, 0)}=p_{j\rightarrow i}^{(0, 1)}=0.5$.  This leads to $p_{j\rightarrow i}^{(0, 1)}+p_{j\rightarrow i}^{(1, 1)}+p_{j\rightarrow i}^{(2, 1)}=p_{j\rightarrow i}^{(0, 1)}$, such that the constraints of node $i$ on all other predecessors are automatically removed. The marginal probability is calculated as

\begin{equation}
p_{i}^{c_{i}}=\frac{e^{-\beta \delta_{ c_{i}}^{0}}(1-\delta_{c_{i}}^{2})\prod\limits_{k\in\partial i^{+}}\sum\limits_{c_{k}\in A^{+}}p_{k\rightarrow i}^{( c_{k}, c_{i})}\prod\limits_{k\in\partial i^{-}}\sum\limits_{c_{k}\in A^{-}}p_{k\rightarrow i}^{( c_{k}, c_{i})}}{\sum\limits_{\acute{ c}_{i}}e^{-\beta \delta_{\acute c_{i}}^{0}}(1-\delta_{c_{i}}^{2})\prod\limits_{k\in\partial i^{+}}\sum\limits_{\acute c_{k}\in A^{+}}p_{k\rightarrow i}^{( \acute c_{k}, \acute c_{i})}\prod\limits_{k\in\partial i^{-}}\sum\limits_{\acute c_{k}\in A^{-}}p_{k\rightarrow i}^{( \acute c_{k}, \acute c_{i})}}.
\end{equation}

If node $i$ is empty but observed (i.e., it has no adjacent occupied parent node but has one occupied 2-distance predecessor), then this node does not restrict the occupation state of any of its unoccupied parent neighbors. For such a node i, the output message $p_{i\rightarrow j}$ or $p_{i\leftarrow j}$ on link $(i,j)$ is then updated according to

\begin{equation}
p_{i\leftarrow j}^{(c_{i},c_{j})}=\frac{e^{-\beta \delta_{ c_{i}}^{0}}A_{i\leftarrow j}^{(c_{i},c_{j})}}{\sum\limits_{\acute{ c}_{i},\acute{ c}_{j}}e^{-\beta \delta_{\acute c_{i}}^{0}}A_{i\leftarrow j}^{(c_{i},c_{j})}},
\end{equation}

where the function $A_{i\leftarrow j}^{(c_{i},c_{j})}$ is defined by equation (4) while $R_{i\leftarrow j}^{(c_{i},c_{j})}$ is redefined as 

\begin{equation}
R_{i\leftarrow j}^{(c_{i},c_{j})}=\delta_{ c_{i}}^{1}(\delta_{ c_{j}}^{c_{i}}+\delta_{ c_{j}}^{c_{i}+1}),
\end{equation}

\begin{equation}
p_{i\rightarrow j}^{(c_{i},c_{j})}=\frac{e^{-\beta \delta_{ c_{i}}^{0}}A_{i\rightarrow j}^{(c_{i},c_{j})}}{\sum\limits_{\acute{ c}_{i},\acute{ c}_{j}}e^{-\beta \delta_{\acute c_{i}}^{0}}A_{i\rightarrow j}^{(\acute c_{i},\acute c_{j})}},
\end{equation}

\begin{equation}
A_{i\rightarrow j}^{(c_{i},c_{j})}=[\prod\limits_{k\in\partial i^{+}}\sum\limits_{c_{k}\in A^{+}}p_{k\rightarrow i}^{( c_{k}, c_{i})}-\delta_{ c_{i}}^{1}\prod\limits_{k\in\partial i^{+}}\sum\limits_{c_{k}\geq c_{i}}p_{k\rightarrow i}^{(c_{k},c_{i})}]\prod\limits_{k\in\partial i^{-}\backslash j}\sum\limits_{c_{k}\in A^{-}}p_{k\rightarrow i}^{( c_{k}, c_{i})}.
\end{equation}

For node $i (c_{i}=2)$, if at least one parent neighbor node $j$ takes the state $c_{j}=1$, then it sends a message to node $i$ as $p_{j\rightarrow i}^{(2, 1)}=p_{j\rightarrow i}^{(2, 2)}=0$.  This leads to $p_{j\rightarrow i}^{(1, 2)}+p_{j\rightarrow i}^{(2, 2)}=p_{j\rightarrow i}^{(1, 2)}$, and the constraint of node $i$ on all other predecessors is automatically removed. The marginal probability is calculated as

\begin{equation}
p_{i}^{c_{i}}=\frac{e^{-\beta \delta_{ c_{i}}^{0}}A_{i}^{c_{i}}}
{\sum\limits_{\acute{ c}_{i}}e^{-\beta \delta_{\acute c_{i}}^{0}}A_{i}^{\acute c_{i}}},
\end{equation}

\begin{equation}
A_{i}^{c_{i}}=[\prod\limits_{k\in\partial i^{+}}\sum\limits_{c_{k}\in A^{+}}p_{k\rightarrow i}^{( c_{k}, c_{i})}-\delta_{ c_{i}}^{1}\prod\limits_{j\in\partial i^{+}}\sum\limits_{c_{j}\geq \acute c_{i}}p_{j\rightarrow i}^{(c_{j},c)}]\prod\limits_{k\in\partial i^{-}}\sum\limits_{c_{k}\in A^{-}}p_{k\rightarrow i}^{( c_{k}, c_{i})}.
\end{equation}

We implement the BPD algorithm as follows.\\
(1) Input network $W$, set all nodes to be unobserved and set all cavity messages $p_{i\rightarrow j}^{(c_{i},c_{j})}$ and $p_{i\leftarrow j}^{(c_{i},c_{j})}$  to be uniform messages. Set the inverse temperature $\beta$ to a sufficiently large value (depending on the highest convergence inverse temperature). Then perform the BP iteration a number T of
rounds (e.g., T = 500). Finally, compute the occupation probability of each node $i$ using Eq. (7).  \\
(2) Cover the small fraction $\gamma$ (e.g., $\gamma=0.01$) of the unfixed nodes having highest covering probabilities.\\
(3) Update the state of all unoccupied nodes: if node $i$ is unoccupied and has at least one parent neighbor that takes state $c_{i}=0$, then it takes state $c_{i}=1$, while if node $i$ is unoccupied and has no occupied parent neighbor, but has at least one parent neighbor that takes state $c_{i}=1$, then it takes state $c_{i}=2$.\\
(4) Fix the state of the observed node; i.e., if the observed node $c_{i}=1$ has at most one successor taking the state $c_{k}=2$, then fix the state of node $i$ as $c_{i}=1$. \\
(5) If network $W$ still contains unobserved nodes, then repeat operations (2)--(4) until all nodes are observed. 

During the decimation process, if the remaining graph still contains unobserved nodes, at least one node is moved to the D-2MDS to reduce the number of unobserved nodes. The BPD process terminates only when no unobserved nodes are present in the remaining graph. We implemented the above BPD algorithm usingthe programming language C++. In our numerical simulations, we set the BPD parameters to be T = 500 and $\gamma= 0.01$. These parameters are not necessarily optimal but are chosen so that a single run of the BPD algorithm on a large graph instance of $N = 10^4$ nodes and $M = 10^5$ edges terminates within several minutes. If the fraction $\gamma$ is further reduced, say to $\gamma= 0.001$, then the BPD algorithm reaches a slightly smaller D-2MDS, but the computing time is much longer. The complexity of the BP algorithm is $O(TNd_{max}^{2})$ or $O(TMd_{max})$ while that of the BPD algorithm is $MAX[O(TNlog(N)),O(TMd_{max})]$, where $d_{max}$ denotes the maximal node degree of the given network.

\subsection{Greedy Algorithm}
We adopted a simple greedy algorithm from the literature to solve the D-2MDS problem approximately. This algorithm is based on the concept of the general impact of a node. The general impact of an unoccupied node $i$ equals the sum of the impact of all successors that are not occupied. The impact of an unoccupied node $i$ equals the number of child nodes that will be observed by occupying node $i$. Starting from an input network W with all nodes unobserved, the greedy algorithm selects uniformly at random a node $i$ from the subset of nodes with the highest general impact and fixes its occupation state to $c_{i}$ = 0. All successors and 2-distance successors of $i$ are then observed. The state of observed nodes is fixed in step (4) of the BPD implementation process. If there are still unobserved nodes in the network, the impact and general impact values for each of the unoccupied nodes are updated and the greedy occupying process is repeated until all nodes are observed. This pure greedy algorithm is easy and fast to implement, and we find that it usually gives results similar to those of the BPD algorithm when the mean arc density of the given network exceeds 20.

\begin{figure}[htb]
  \centering
  \includegraphics[width=12cm,height=7cm]{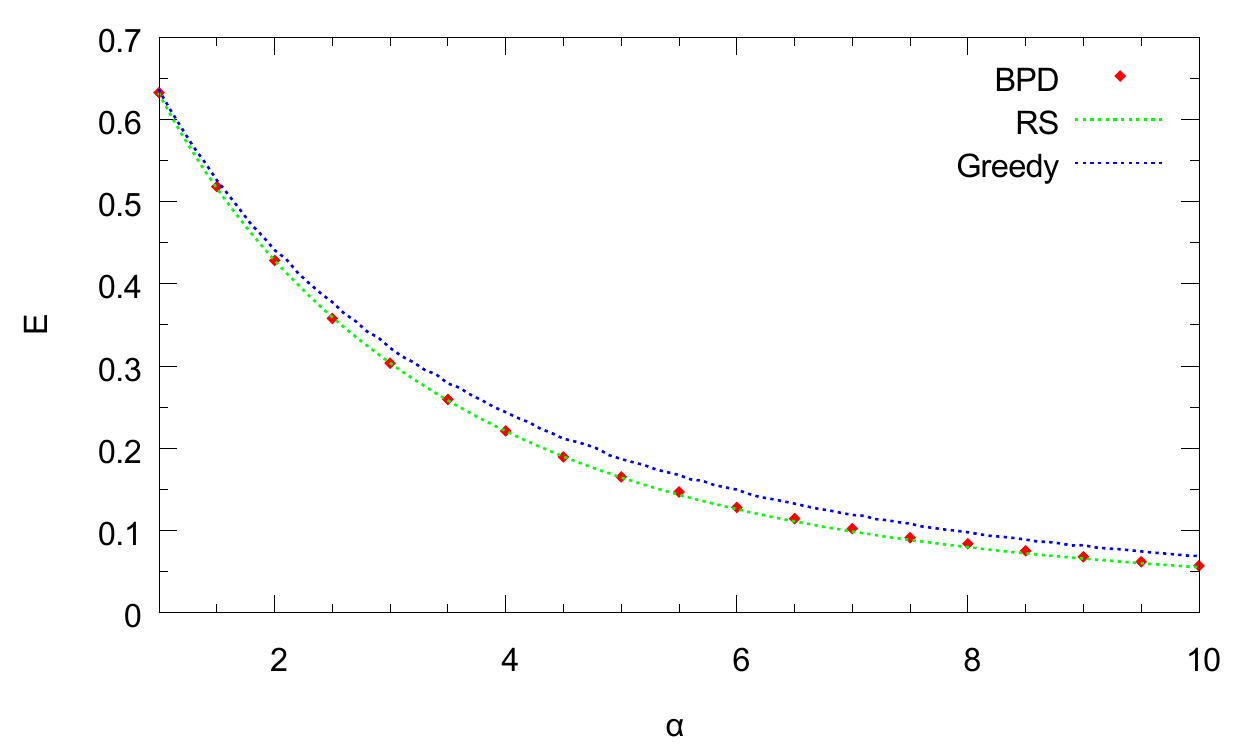}
  \caption{ BPD, RS and greedy algorithm results for the D-2MDS problem on the ER random graph with a size of $N=10^{4}$ nodes. The BPD and greedy algorithm results are obtained on five ER random graphs that include $N=10^{4}$ nodes. The $x$-axis denotes the mean arc density while and $y$-axis denotes the energy density. Inverse temperature $\beta = 10.0$.}
\end{figure}
\begin{figure}[htb]
  \centering
  \includegraphics[width=12cm,height=7cm]{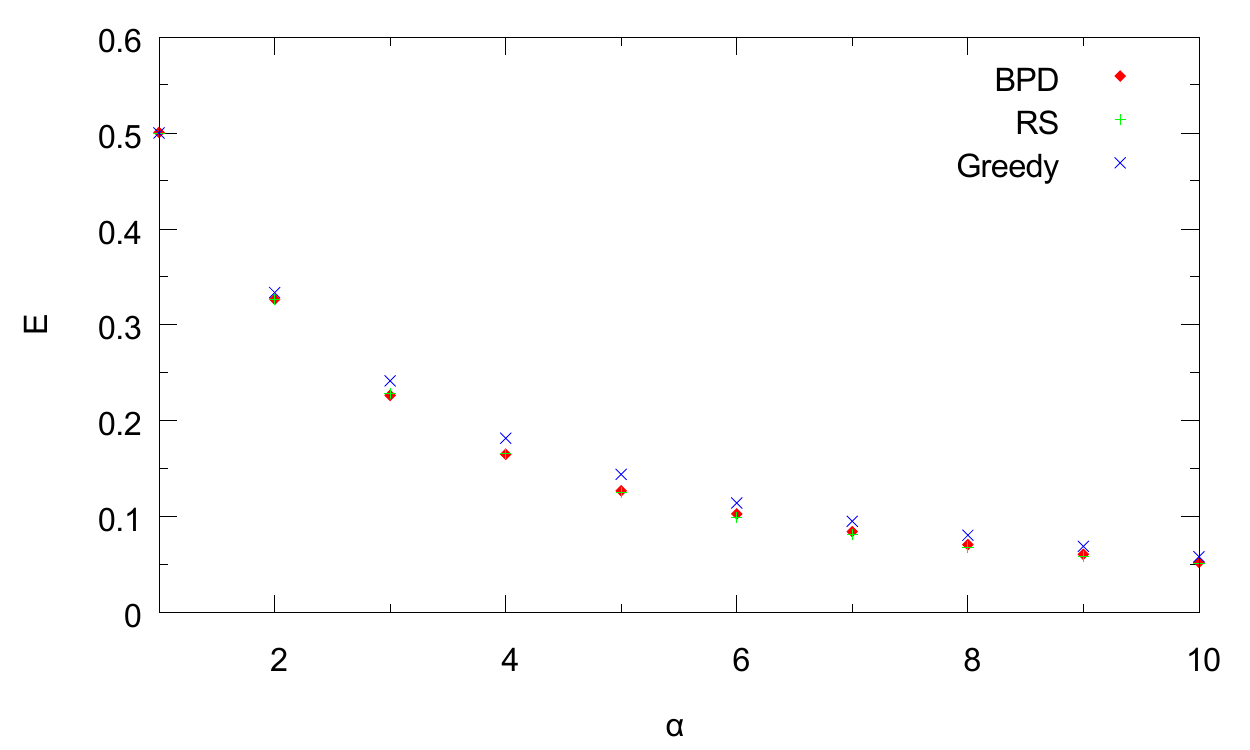}
  \caption{ BPD, Greedy and RS results for the D-2MDS problem on the RR random graph with a size of $N=10^{4}$ nodes. The $x$-axis denotes the mean arc density while the $y$-axis denotes the energy density. Inverse temperature $\beta = 10.0$.}
\end{figure}

The results of the greedy algorithm for the ER random network and RR network are compared with the results of the BPD algorithm in Figs. 4 and 5. The BPD algorithm outperforms the greedy algorithm and provides results that are close to those of RS theory on both the ER random and RR graphs.
\section{Discussion}
We proposed two algorithms (i.e., a greedy-impact local algorithm and a BPD message-passing algorithm) and presented an RS mean field theory for solving the D-2MDS problem algorithmically and theoretically. We found that the BP and RS algorithms lead to an entropy transition (from a positive value to a negative value) on both ER and RR networks when the mean arc density exceeds a threshold but there is no entropy transition when the mean arc density is lower than this threshold value (i.e., a value of 4 for the RR network and 6.6 for the ER random network). The reason for this result is that the solution space of the D-2MDS problem on the two networks has a structural transition. We will use one-step RS breaking theory to study the solution space of the D-2MDS problem. Our numerical results shown in Figs. 4 and 5 suggest that the mean-field BPD algorithm constructs a near-optimal D-2MDS for random networks and the mean-field BPD algorithm is better than the greedy algorithm.\\
There is much theoretical work remaining to be done. We will soon work on the one-step RS breaking of the D-2MDS problem. A more challenging and common problem for the dominating set is the directed connected dominating set problem, and we will use spin glass theory [25] to study the directed minimally connected dominating set problem and the directed 2-distance minimally connected dominating set problem.
\section{$\hspace{2mm}$ Acknowledgement}
This research was supported by the doctoral startup fund of Xinjiang University of China (grant number 208-61357) and the National Natural Science Foundation of China (grant number 11465019, 11664040). We thank Glenn Pennycook, MSc, from Liwen Bianji, Edanz Group China (www.liwenbianji.cn/ac), for editing the English text of a draft of this manuscript.

\end{document}